\documentclass[review, numbers,sort&compress]{elsarticle}

\usepackage{lineno,hyperref}
\modulolinenumbers[5]

\journal{Physics Letters B}

\begin{document}

\begin{frontmatter}

\title{Neutron skin and signature of the $N$ = 14 shell gap found from measured proton radii of $^{17-22}$N}

\author[label1,label2,label3]{S. Bagchi}
\author[label1,label4]{R. Kanungo} 
\author[label5]{W. Horiuchi}  
\author[label6,label7]{G. Hagen\fnref{fn1}}
\author[label6,label7]{T.~D.~Morris\fnref{fn1}}
\author[label4]{S. R. Stroberg} 
\author[label8]{T. Suzuki} 
\author[label2]{F. Ameil} 
\author[label1]{J. Atkinson} 
\author[label9]{Y. Ayyad} 
\author[label9]{D. Cortina-Gil}
\author[label2]{I. Dillmann\fnref{fn2}}
\author[label1,label2]{A. Estrad\'e\fnref{fn3}}
\author[label2]{A. Evdokimov}
\author[label2]{F. Farinon}
\author[label2,label3]{H. Geissel}
\author[label2]{G. Guastalla}
\author[label10]{R. Janik}
\author[label1,label11]{S. Kaur}
\author[label2]{R. Kn\"obel}
\author[label2]{J. Kurcewicz}
\author[label2]{Yu. A. Litvinov}
\author[label2]{M. Marta}
\author[label9]{M. Mostazo}
\author[label2]{I. Mukha}
\author[label2]{C. Nociforo}
\author[label12]{H.J. Ong}
\author[label2]{S. Pietri}
\author[label2]{A. Prochazka}
\author[label2,label3]{C. Scheidenberger}
\author[label10]{B. Sitar}
\author[label10]{P. Strmen}
\author[label2]{M. Takechi\fnref{fn4}}
\author[label12]{J. Tanaka}
\author[label1,label2,label3]{Y. Tanaka}
\author[label12,label13]{I. Tanihata}
\author[label13]{S. Terashima}
\author[label9]{J. Vargas}
\author[label2]{H. Weick}
\author[label2]{J. S. Winfield}

\fntext[fn1]{This manuscript has been authored by UT-Battelle, LLC under Contract No. DE-AC05-00OR22725 with the U.S. Department of Energy. The United States Government retains and the publisher, by accepting the article for publication, acknowledges that the United States Government retains a non-exclusive, paid-up, irrevocable, world-wide license to publish or reproduce the published form of this manuscript, or allow others to do so, for United States Government purposes. The Department of Energy will provide
public access to these results of federally sponsored research in accordance with the DOE Public Access Plan
(http://energy.gov/downloads/doe-public-access-plan)}
\fntext[fn2]{Present address : TRIUMF, Vancouver, BC V6T 4A3, Canada}
\fntext[fn3]{Present address : Department of Physics, Central Michigan University, Mount Pleasant, MI 48859, USA}
\fntext[fn4]{Present address : Department of Physics, Niigata University, Niigata 950-2181, Japan}
\address[label1]{Astronomy and Physics Department, Saint Mary's University, Halifax, NS B3H 3C3, Canada}
\address[label2]{GSI Helmholtzzentrum f\"ur Schwerionenforschung GmbH, D-64291 Darmstadt, Germany}
\address[label3]{Justus-Liebig University,  35392 Giessen, Germany}
\address[label4]{TRIUMF, Vancouver, BC V6T 4A3, Canada}
\address[label5]{Department of Physics, Hokkaido University, Sapporo 060-0810, Japan}
\address[label6]{Physics Division, Oak Ridge National Laboratory, Oak Ridge, TN 37831, USA}
\address[label7]{Department of Physics and Astronomy, University of Tennessee, Knoxville, TN 37996, USA}
\address[label8]{Department of Physics, Nihon University, Setagaya-ku, Tokyo 156-8550, Japan}
\address[label9]{Universidad de Santiago de Compostela, E-15706 Santiago de Compostella, Spain }
\address[label10]{Faculty of Mathematics and Physics, Comenius University, 84215 Bratislava, Slovakia}
\address[label11]{Department of Physics and Atmospheric Science, Dalhousie University, Halifax, NS B3H 4R2, Canada}
\address[label12]{RCNP, Osaka University, Mihogaoka, Ibaraki, Osaka 567 0047, Japan}
\address[label13]{School of Physics and Nuclear Energy Engineering and IRCNPC, Beihang University, Beijing 100191, China}

\begin{abstract}
A thick neutron skin emerges from the first determination of root mean square radii of the proton distributions for $^{17-22}$N from charge changing cross section measurements around 900$A$ MeV at GSI.  Neutron halo effects are signalled for $^{22}$N from an increase in the proton and matter radii. The radii suggest an unconventional shell gap at  $N$ = 14 arising from the attractive proton-neutron tensor interaction, in good agreement with shell model calculations. {\it Ab initio}, in-medium similarity re-normalization group, calculations with a state-of-the-art chiral nucleon-nucleon and three-nucleon interaction reproduce well the data approaching the neutron drip-line isotopes but are challenged in explaining the complete isotopic trend of the radii. 
\end{abstract}


\end{frontmatter}

Neutron-rich nuclei are fertile grounds to search for unexpected features. Exotic nuclear forms are
revealed with the formation of neutron skins and halos approaching the neutron drip-line \cite{TA85,HA87,TA13}, many of which
relate to modifications of conventional shells. The emerging signatures of changes in the 
shell structure must be identified and their origins understood. 
The presence of neutron halos in $^{11}$Li and $^{11}$Be relate to the breakdown of the $N$ = 8 shell gap.
Evidence has been found for a new shell gap at $N$ = 16 at the drip-line of carbon to fluorine isotopes. 
Studies of excited states and momentum distributions have discussed a shell gap at $N$ = 14 between the 1$d_{5/2}$ and 2$s_{1/2}$ orbitals in oxygen isotopes.
However, its reduction for the nitrogen isotopes is signalled and its disappearance due to a level inversion is predicted in the carbon isotopes \cite{RO11,SO08,BE01,ST04,BE06,EL10,DI18,SC07,ST09}. 
Therefore, further experimental investigation is needed for revealing the cause of this shell gap and its evolution. 
Systematic trends of proton radii along an isotopic chain can reveal the presence of neutron magic numbers \cite{AN13}.  

In this work the first determination of root mean square radii of density distribution of protons treated as point particles, referred to henceforth as point proton radii of neutron-rich isotopes $^{17-22}$N together with those for stable nuclei $^{14,15}$N is presented from a measurement of charge changing cross
sections. 
The proton radii decrease from $^{17}$N to $^{21}$N with this hint of a minimum reflecting a shell gap at $N$ = 14 in the nitrogen isotopes. The proton radius increases beyond this for $^{22}$N. In a $^{21}$N (core) + $n$ model of $^{22}$N, this shows an enlargement of the core $^{21}$N. The evolution of matter and proton radii reveal thick neutron skins in $^{18-22}$N.  

One neutron removal reactions show a reduction in the width of the longitudinal momentum distribution ($P_{||}$) between $^{21}$N ($\Delta P_{||}  = $160$\pm$32 MeV/c) and $^{22}$N ($\Delta P_{||} =$ 77$\pm$32 MeV/c) that indicates a change of dominating neutron orbitals from $l = $ 2 to $l =$ 0 \cite{RO11}. 
The $P_{||}$ for $^{18,19}$N are explained by  $\sim$ 69\% probability of the neutron in the $l =$ 2 orbital with the core nucleus in excited states. 
The situation changes in $^{20,21}$N  where the $P_{||}$ are explained with 83\% and 68\% probability, respectively of
valence neutrons in the $l =$ 2 orbital with the core in its ground state. A shell gap at $N =$ 14 in $^{22}$O was first indicated from the high excitation energy of its first excited state \cite{BE01,ST04}. Proton inelastic scattering \cite{BE06} affirms this, implying a small quadrupole deformation ($\beta$ = 0.26(4)) and a $B(E2)$ value deduced to be 21(8) e$^2$fm$^4$.  A measure of the $N =$ 14 energy gap in $^{23}$O was derived to be 2.79(13) MeV from the observation of an unbound 1$d_{5/2}$ hole state \cite{SC07}. However, proton inelastic scattering of $^{21}$N is consistent with a much larger $B(E2)$ value of 56(18) e$^2$fm$^4$  \cite{EL10} and a reduction of the $N =$ 14 shell gap by 1.2 MeV in going from O to N. Recently, quasifree knockout reaction studies of $^{22,23}$O and $^{21}$N show a decrease in the width of the momentum distribution in going from $^{22}$O to $^{21}$N suggesting a reduced $N =$ 14 shell gap for this nucleus leading to more configuration mixing of the 2$s_{1/2}$ orbital \cite{DI18}.  This shell gap was found to be strongly reduced to 1.41(17) MeV in $^{22}$N and predicted to disappear in the carbon isotopes \cite{ST09}.  Ref.\cite{SO08} however deduces a moderately large energy gap of 3.02 MeV at $N =$ 14 from the excited states in $^{21}$N. Therefore, more experimental information is needed to understand if $N =$ 14 is a shell gap in the nitrogen nuclei which is at the transition region between the oxygen and carbon nuclei. 

Point proton radii of nuclei reflect deformation and shell effects. Shell gaps can be visible as local minima in these radii along an isotopic chain \cite{AN13} and can bring new insight into shell evolution. The charge radii of $^{18-28}$Ne \cite{MA11}, from isotope shift measurements show local minima at $N = $8 and 14. 

\begin{figure}
\includegraphics[width=9cm, height=6.5cm]{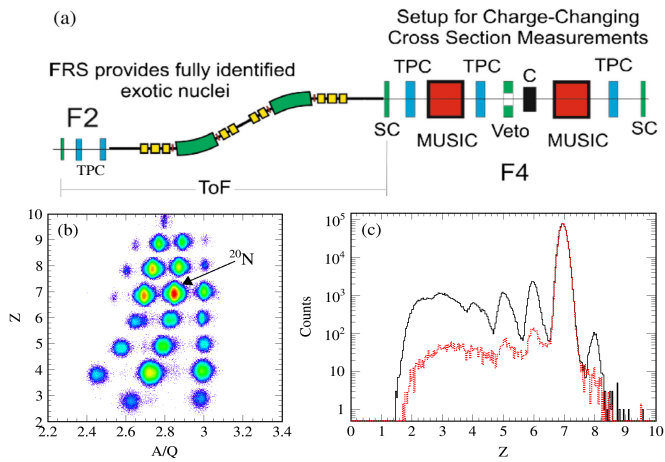}
\caption{(Color online) (a) Schematic view of the second half of the FRS spectrometer with the detector arrangements at the intermediate focal plane F2 and at the final focal plane F4. (b) Particle identification spectrum before the C reaction target with $^{20}$N indicated by an arrow. (c) $Z$ spectrum in MUSIC after the reaction target with $^{20}$N events selected before the target. The black solid and red dotted histograms are with and without the reaction target, respectively, with the latter normalized by factor 1.4 to the $Z$~=~7 peak of the former.}
\label{Fig1}
\end{figure}

The first determination of point proton radii ($R_p$) from measurements of charge-changing cross sections ($\sigma_{cc}$) for the neutron-rich isotopes $^{17-22}$N as well as for the stable isotopes $^{14,15}$N are reported here. The experiment was performed at the fragment separator FRS \cite{FRS} at GSI in Germany. Beams of $^{14,15}$N and $^{17-22}$N were produced by fragmentation of $^{22}$Ne and $^{40}$Ar, respectively, at 1$A$ GeV, on a 6.3 g/cm$^2$ thick Be target. The isotopes of interest were separated in flight and identified using their magnetic rigidity ($B\rho$), time-of-flight and the energy-loss measured in a multi-sampling ionization chamber (MUSIC) \cite{MUSIC}. In Fig. \ref{Fig1}(a), the second half of the FRS is shown along with the detector configuration at the final achromatic focus of the separator. The time-of-flight of the incoming beam was measured from the dispersive intermediate focal plane F2 to the achromatic final focal plane F4 using plastic scintillators. 
Beam tracking with position sensitive time-projection chambers (TPC) \cite{TPC} placed at F2 and at F4 provided event-by-event $B\rho$ determination of the incoming particles. A  4.010 g/cm$^2$ thick carbon reaction target was placed at F4. The total beam rate at F4 varied from 1600-4000 / spill for the different fragment settings. The rates of $^{14,15,17-22}$N were $\sim$ 70, 6, 100, 15, 100, 90, 80, 2 pps, respectively.

The basic principle of obtaining $\sigma_{cc}$ is the transmission technique, where the number ($N_{\rm{in}}$) of incident nuclei of interest $^{A}Z$ is determined from event-by-event counting. After the reaction target, the number of nuclei with the same charge $Z$ as the incident nuclei ($N_{\rm{SameZ}}$) are identified using a MUSIC detector. The $\sigma_{cc}$ is then obtained from the relation $\sigma_{cc}$ = $t^{-1}ln$($R_{\rm{T}_{out}}$/$R_{\rm{T}_{in}}$). Here $R_{\rm{T}_{in}}$ and $R_{\rm{T}_{out}}$ are the ratios of $N_{\rm{SameZ}}$/$N_{\rm{in}}$ with and without the reaction target, respectively and $t$ is the target thickness. The term $R_{\rm{T}_{out}}$ accounts for losses due to interactions with non-target materials and for detection efficiencies. 

A restriction on the phase space of the incident beam at the target eliminated events having large incident angles and position far from the center. A veto scintillator with a central aperture placed in front of the target gave the possibility of rejecting beam events incident on the edges of the target scattered by matter upstream and multi-hit events that can cause erroneous reaction information in the MUSIC placed after the target. In the selection of nitrogen nuclei, the admixture of $Z$ = 6 and 8 relative to $Z$ = 7 is $<$ 10$^{-4}$. Fig.~\ref{Fig1}(b) shows a particle identification spectrum before the target where the black circled events denote the isotope of interest and the other events are contaminant fragments. 

With the desired nitrogen isotope events ($N_{in}$) selected for the incident beam, the events having $Z$ = 7 after the reaction target were counted using the MUSIC detector placed downstream of the target. This yielded $N_{\rm{SameZ}}$. The TPC and the plastic scintillator detectors placed downstream of the second MUSIC provided additional $Z$ information and their correlation with the MUSIC detector ensured proper $Z$ identification and MUSIC efficiency determination. The 1$\sigma$ $Z$ resolution for nitrogen in the MUSIC was $\Delta Z$$\sim$0.11. To obtain $N_{\rm{SameZ}}$, a selection window of width $\sim $4$\sigma$ was put around the $Z$~=~7 peak (Fig.~\ref{Fig1}(c)) in the low-$Z$ side of the spectrum. On the high-$Z$ side, the selection window includes the peak at $Z$~=~8  (Fig. 1(c)). 
This is because an increase in $Z$ does not arise from interactions between protons in the incident N isotope and target nucleons. Therefore, they need to be included in the unchanged charge events ($N_{\rm{SameZ}}$) used to measure $\sigma_{cc}$ for extracting $R_p$.           

For the stable nucleus $^{14}$N the measured $\sigma_{cc}$ is 828$\pm$5 mb. However, one-neutron removal here can lead to $^{13}$N, in proton-unbound excited states since the one-proton separation energy is very low ($S_p$ = 1.9 MeV). In such a situation decay by proton emission changes $Z$, although this process does not involve any interaction with the protons.  The corrected $\sigma_{cc}$ for $^{14}$N is therefore obtained by subtracting the one-neutron removal cross section that leads to proton unbound states in $^{13}$N. This cross section is estimated to be 35$\pm$7 mb, using the Glauber model and spectroscopic factors based on pickup reaction measurements \cite{HI67} as well as from shell model calculations using the WBP Hamiltonian in
the $p$-shell \cite{Brpr}.  For $^{15}$N, the one-neutron removal cross section to $^{14}$N states above the proton threshold is estimated to be 12$\pm$5 mb using spectroscopic factors from
Ref.\cite{Sn69}.  The measured cross section of 828$\pm$20 mb for $^{15}$N is therefore corrected to eliminate this neutron removal effect. The corrected $\sigma_{cc}$ for $^{14,15}$N are listed in Table 1. For neutron-rich nuclei $^{17-22}$N the proton separation energy gradually increases thereby greatly decreasing the neutron removal cross section to proton unbound states. Hence, no correction of $\sigma_{cc}$ is necessary for these nuclei. 

The measured cross sections with one standard deviation total uncertainties are listed in Table 1.  The uncertainties contain a $\sim$0.1\% contribution from target thickness,  uncertainties from contaminants in the beam events and fluctuations within the selected phase space.  The $\sigma_{cc}$ values decrease with lowering of the event rejection threshold of the veto detector. Table 1 shows the $\sigma_{cc}$ values with complete rejection of events hitting the veto detector and no rejection of veto hit events. 
It may be mentioned that the $\sigma_{cc}$ reported in Ref.\cite{CH00} have large uncertainties
thereby making them unsuitable for discussing nuclear structure evolution and neutron skin thickness. No radii were determined in Ref.\cite{CH00}. 

The cross sections are analyzed within the Glauber model framework described in Ref.\cite{SU16} and 
used in Refs.\cite{ES14,TE14,KA16}  to derive the point proton radii. A harmonic oscillator point proton density distribution is considered without any recoil effect from the neutron distribution. 
The reaction cross section ($\sigma_{R}$) is calculated with the nucleon-target formalism in Glauber theory \cite{AB00} which effectively includes the multiple-scattering effect missing in the optical-limit approximation \cite{AB00,HO07}. At the high beam energies of this experiment, the interaction cross section ($\sigma_{int}$) is approximately equal to  $\sigma_{R}$. Both the $\sigma_{cc}$  and $\sigma_{R}$ are evaluated with the finite-range profile function parametrized in Ref.\cite{AB08}. Once the input
densities for the projectile and target, and the profile function parameters are fixed, the theory has no adjustable parameters. A study of $\sigma_{R}$ and $\sigma_{cc}$ for $^{12}$C+$^{12}$C in Ref.\cite{SU16} shows that the uncertainty from profile function parameters at energies around 900$A$ MeV is less than 1\% from the consistency of finite range calculations with known density of $^{12}$C. This is also seen from the comparison of proton radii for $^{12,13}$C from $\sigma_{cc}$ and e$^-$ scattering \cite{KA16}. The systematic uncertainty in $\sigma_{R}$ from different projectile densities with the same radius is around 5\% \cite{HO14} mainly due to differences in the neutron density tail. Ref.\cite{KA11} shows that a 2\% uncertainty in $\sigma_{R}$ leads to 5\% uncertainty in the matter radius with different densities. This uncertainty is smaller for the proton radius since the density of the deeply bound protons does not have an extended tail and the $\sigma_{cc}$ measured have $\sim$0.6-1\% uncertainty. The harmonic oscillator density profile used here is well justified and therefore no significant systematic uncertainty can be foreseen in the radii reported here. Systematic uncertainties however, do not change the relative isotopic variation of radii. 

Using the $\sigma_{cc}$ with full veto rejection, the point proton radius R$_p^{ex,veto}$ of $^{14}$N derived from $\sigma_{cc}$ is consistent with that (R$_p^{(e^-)}$) from electron scattering \cite{AN13} (Table 1). The R$_p^{ex}$ for $^{10}$B and $^{12-14}$C \cite{ES14,KA16} from  $\sigma_{cc}$ without any scaling normalization have also been found to be consistent with point proton radii derived from electron scattering. With increasing rejection threshold of the veto detector (i.e. less rejection) the $\sigma_{cc}$ value increases. The ratio of R$_p$ of $^{14}$N for increasing veto thresholds to R$_p^{(e^-)}$ was found to increase from 1.00(01) to 1.07(01). This ratio factor was used to scale the R$_p^{ex}$ of $^{15-22}$N for the different thresholds. The average of the radii from seven different thresholds (from full rejection, i.e. rejecting pulse height above pedestal, up to no rejection, i.e. including pulse height with ADC overflow) after scaling is listed in Table 1 as R$_p^{ex,avg}$. 
It is seen that though consistent within uncertainties for some isotopes central values of the R$_p^{ex,avg}$ are slightly greater than R$_p^{ex,veto}$ which was found to arise largely from large pulse height overflow events in the veto detector. 

\begin{table}
\caption{\label{tab:table1} Secondary beam energies, measured $\sigma_{cc}$ and the root mean square point proton and matter radii derived for the nitrogen isotopes.(\it{See text for symbols}).}
\begin{tabular}{llllllll}
Isotope & E/A &$\sigma_{cc}^{ex,veto}$&$\sigma_{cc}^{ex,no veto}$&$R_p^{ex,veto}$&$R_p^{ex,avg}$&$R_p^{(e^-)}$ &$R_m^{ex}$\\
&(MeV)&(mb)&(mb)&(fm)&(fm)&(fm)&(fm)\\
\hline
$^{14}$N& 932  & 793(9)&833(9)&2.43(4) &2.43(4)&2.43(1) &2.50(3) \\
$^{15}$N& 776 & 816(20)&843(20)&2.55(9)&2.49(9)&2.49(1) &2.44(10)\\
$^{17}$N& 938 & 819(5)&874(5)&2.54(2)&2.55(3)& &2.52(8)  \\
$^{18}$N& 927 & 810(6)&869(6)&2.51(2)&2.53(3)& & 2.68(2) \\
$^{19}$N& 896 & 809(5)&864(5)&2.51(2)&2.52(3)& &2.74(3) \\
$^{20}$N& 891 & 808(5)&866(5)&2.50(2)&2.52(3)&&2.84(5)\\
$^{21}$N& 876 & 799(7)&857(7)&2.46(3)&2.49(3)& &2.78(2)\\
$^{22}$N& 851 & 810(7)&869(7)&2.51(3)&2.53(3)& &3.08(12)\\
\end{tabular}
\end{table}

The extracted point proton radii $(R_p^{ex})$ from $\sigma_{cc}$ are shown in Fig.2 with black filled circles for R$_p^{ex,veto}$, black open circles for the average proton radii (R$_p^{ex,avg}$) extracted for different thresholds of the veto detector and those from electron scattering for $^{14,15}$N are shown by the black stars. A decrease in $R_p^{ex,avg}$ from $^{17}$N to $^{21}$N  is seen from the best fit line (black line in Fig.2a) to the data including the uncertainties whose slope is -0.0130 $\pm$ 0.0095 fm/$A$. Beyond $^{21}$N an increase in $R_p^{ex,avg}$ is observed for $^{22}$N. 
To test a null hypothesis we performed the statistical $F$-test with a constant radius fit for $^{17}$N to $^{22}$N and the linear fit discussed above using the respective reduced chisquare value. For this comparison an $F$ value of 6.11 obtained with the relevant degrees of freedom, 5 and 3,  shows that the data exhibit a dip in the proton radius of $^{21}$N with $p$- value of 93\% probability. The decrease in the central value of the proton radius for $^{21}$N compared to $^{20,22}$N is found even with lower beam counts, providing support that this feature is not arising from statistical fluctuations. The statistical contribution to the uncertainty of the proton radii for $^{20,21}$N is $\sim$ 0.01 fm, and for $^{22}$N is $\sim$ 0.03 fm. 

The decrease of $R_p^{ex,avg}$  with filling of the 1$d_{5/2}$ orbital reflects the strong attractive interaction between the proton (1$p_{1/2}$) and the neutrons (1$d_{5/2}$) leading to the emergence of a shell gap at $N$ = 14 and a reduction in deformation. An increase in $R_p^{ex,avg}$ observed, within the uncertainty, for $^{22}$N results from its extended neutron density for the valence neutron in the 2$s_{1/2}$ orbital with a closed-shell core of $^{21}$N. This causes the center-of-mass ($c.m.$) of $^{22}$N to be different from that of the core $^{21}$N leading to $c.m.$ motion smearing of the core density and hence a larger core size.  The extended core root mean square matter radius, following Ref.\cite{TA13}, is found to be 2.82 fm considering the central radius values.
The valence neutron wavefunction in $^{22}$N is calculated by solving the Schr\"odinger equation with a Woods-Saxon binding potential where the potential depth is adjusted to reproduce the one-neutron separation energy of 1.28 MeV \cite{AME12}. Using this wavefunction and the extended $^{21}$N core size the $\sigma_{R}$ of $^{22}$N + $^{12}$C is calculated in a core + neutron few body Glauber model. The resulting $\sigma_{R}$ is 1221 mb which is in good agreement with the measured cross section of 1245$\pm$49 mb \cite{OZ01}. Considering an admixture of the 1$d_{5/2}$ neutron orbital with the $^{21}$N core in the 3/2$^-$, 1.16 MeV excited state the data are consistent with a valence neutron probability of 50\% - 100\% in the 2$s_{1/2}$ orbital. This agrees within the upper end of the uncertainty band with the momentum distribution measurement \cite{RO11}.  Therefore the evolution of  $R_p^{ex}$ sheds new light on the halo-like structure in $^{22}$N. The $R_p^{ex}$  of $^{15,18-22}$N are consistent with the relativistic mean field (RMF) predictions of Ref. \cite{ME02, ZH96} (Fig.2b) while the radii predicted in Ref.\cite{LA04} are larger than the data. 

\begin{figure}
\includegraphics[width=10cm, height=5cm]{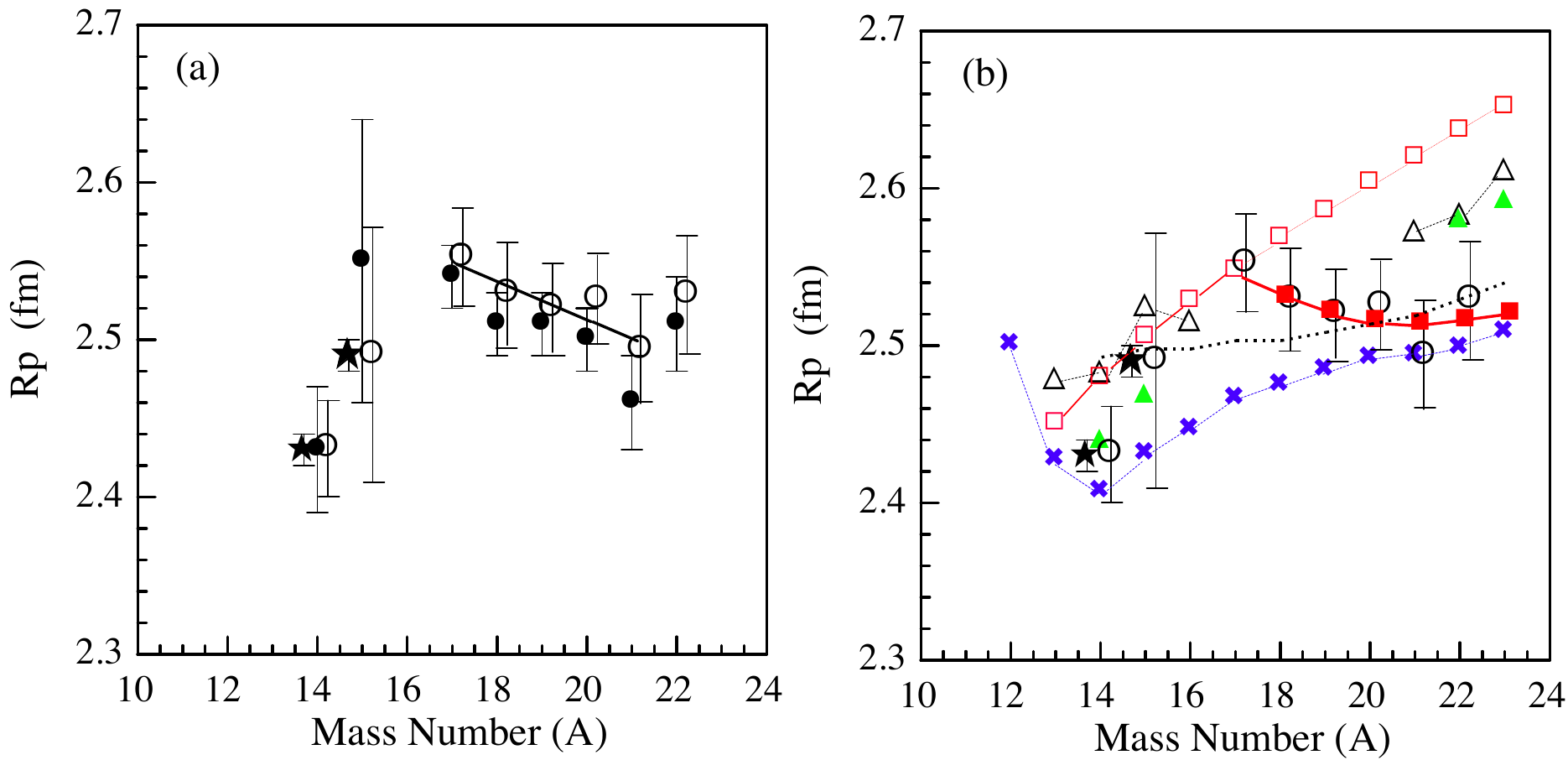}
\caption{\label{fig:epsart} (Color online) (a) The experimental point proton radii (black filled circles are $R_p^{ex,veto}$ and black open circles are $R_p^{ex,avg}$, stars from $e^-$ scattering).  The black solid line shows the best linear fit including the uncertainties. (b) The open circles are $R_p^{ex,avg}$ compared to theoretical predictions. Shell model predictions using the YSOX interaction and WS (HO) potentials are shown with red filled (open) squares and the VS-IMSRG radii are indicated with blue crosses. The red and blue lines are guides to the eye. The coupled-cluster radii computed with carbon cores and oxygen cores are shown with green filled triangles and black open triangles, respectively. Relativistic mean field theory predictions are shown by the black dotted curve from Ref.\cite{ME02}.}
\end{figure}

The proton radii determined are compared to shell model predictions and {\it ab initio} calculations in Fig.2b. The shell model evaluations use a harmonic oscillator (HO) basis for nuclei close to stability, $^{13-17}$N, while Woods-Saxon (WS) wave functions are used for the neutron-rich isotopes, $^{18-23}$N. The evaluation is carried out by using the proton occupation numbers of $p$, $s$ and $d$ orbitals for each isotope obtained with the YSOX Hamiltonian \cite{YSOX} including the tensor force, that reproduces  the ground-state energies, drip lines, energy levels, electric quadrupole properties and spin properties of B to O isotopes. 
 HO wave functions obtained with $\hbar\omega$ = (45/$A^{1/3}$$-$25/$A^{2/3}$)  MeV yield proton radii that increase as $A^{1/6}$  since the potential width becomes larger with increasing mass number $A$ (red open squares in Fig.2b shown for comparison). However, as $N$ increases, the depth of the one-body potential for proton must increase due to more attractive neutron-proton interaction. This effect is therefore, taken into account for neutron-rich isotopes by adopting a WS potential, whose strengths of the central and spin-orbit parts are taken to be $V_{0}$ = $-$51.0 $-$33 $(N-Z)/A$ MeV and $V_{LS}$ = 0.44$\times V_{0}$ MeV, respectively \cite{BM}. The radius parameter is 1.27 fm and the diffuseness 0.67 fm.  The combination of HO and WS wavefunctions show that the proton radii of the nitrogen isotopes increase for $A$ = 13-17, beyond which a decrease is seen leading to a shallow minimum at $N$ = 14 and a very small increase beyond it.  The predictions reproduce the experimental data rather well especially for $^{17-22}$N. 

{\it Ab initio} calculations were performed with the chiral nucleon-nucleon and three-nucleon potential N$^2$LO$_{\rm{sat}}$ \cite{ekstrom2015}, which has been shown to describe binding energies and radii of light and medium-mass nuclei accurately \cite{ekstrom2015,hagen2015,garciaruiz2016,duguet2017}. We employ two different theoretical frameworks, namely the coupled-cluster~\cite{bartlett2007,hagen2014} and valence-space in-medium similarity renormalization group (VS-IMSRG) methods~\cite{tsukiyama2012,bogner2014, stroberg2016,stroberg2017}.  Coupled-cluster calculations use a Hartree-Fock basis of 15 major oscillator shells with $\hbar\omega = 16$~MeV, while VS-IMSRG use a basis of 11 major oscillator shells with $\hbar\omega = 22$~MeV. In our coupled-cluster calculations we apply the normal-ordered two-body approximation \cite{hagen2007a,roth2012,binder2013b} for the $NNN$ interaction, with an additional energy cut on three-body matrix elements $e_1+e_2+e_3\leq E_{\rm 3 max} = 16$ $\hbar\omega$. The VS-IMSRG calculations use a Hartree-Fock like reference which is constructed with respect to ensemble states above $^{4}$He for $^{14-15}$N, and states above $^{10}$He for $^{16-23}$N following Ref.~\cite{stroberg2017}.  This allows for a similar normal-ordered two-body approximation for the $NNN$ interaction, with $E_{\rm 3 max} = 14$ $\hbar\omega$. The coupled-cluster calculations of 
$^{13-17}$N and $^{21-23}$N are done by employing generalized equation-of-motion states obtained by charge-exchange, particle-removed, and particle-attached from closed shell calculations of carbon and oxygen isotopes~\cite{gour2005,gour2008,jansen2011,jansen2012,ekstrom2014,hagen2014}.  Charge-exchange, particle-removed, and particle-attached calculations are approximated at the coupled-cluster singles and doubles, $2p-1h$, and $1p-2h$ excitation level, respectively. In addition, we augment the  particle-attached (removed) calculations with the newly developed $3p-2h$ ($2p-3h$) perturbative corrections~\cite{morris2017}. These are in essence based on the completely renormalized coupled-cluster formalism~\cite{kowalski2000,piecuch2002,binder2013}. On the other hand, the VS-IMSRG approach generates an effective shell-model interaction which can be diagonalized by conventional means such that all nitrogen isotopes within the valence space can be calculated. 

Coupled-cluster calculations for point-proton radii starting from different closed (sub-) shell $^{14,16,22,24}$O and $^{14,22}$C isotopes, are shown in Fig.~2b by black open triangles and green filled triangles, respectively. The calculated radii of $^{14,15}$N starting from either $^{14}$C or $^{14,16}$O disagree by about~0.05 fm.
The disagreement in $^{14,15}$N gives us an estimate of errors in the employed coupled-cluster truncations. VS-IMSRG results for point-proton radii are also shown by blue crosses (Fig. 2b), and are in agreement with the measured radii for $^{19-22}$N and just slightly lower than the data for $^{14}$N. 
The decrease in $R_p^{ex}$ observed experimentally for $^{17-21}$N is not successfully predicted by either of the {\it ab initio} frameworks. An indication of a flattening in $R_p$ for $N$ = 13-15 is seen in VS-IMSRG calculations, which predict a prominent dip in radius for $^{14}$N. We note that {\it ab initio} calculations of neutron-rich calcium isotopes yielded a similar discrepancy with observed radii \cite{garciaruiz2016}.

The matter radii ($R_{m}^{ex}$) of $^{14,15,17-22}$N are extracted in this work using the Glauber model where the proton radius of R$_p^{ex,avg}$ is used and the neutron radius from harmonic oscillator density is varied to get the different matter radii that reproduce the measured $\sigma_{int}$ \cite{OZ01}.  For $^{17}$N,  the $\sigma_{int}$ data at 710$A$ MeV \cite{OZ01} are adopted for $R_m^{ex}$ and neutron skin. The extracted $R_{m}^{ex}$, listed in Table 1 and shown in Fig. 3(a) (black open circles), rapidly increases from $^{18-22}$N showing a small dip at the $N$ = 14 nucleus $^{21}$N. The coupled-cluster and VS-IMSRG calculations does not reproduce this dip in $R_{m}^{ex}$ from $^{21}$N to $^{22}$N, but are in good agreement with the magnitude for the other isotopes. Fig. 3(b) shows the neutron skin thickness, $R_n$ - $R_p$ linearly increasing for $^{18-22}$N with a small decrease at $N$ = 14. The {\it ab initio} predictions are consistent with this rapid increase in skin thickness for $^{19-22}$N approaching the neutron drip-line.  

The reason for appearance of a dip in $R_p$ at $N =$ 14 in the Ne and N isotopes is different. In Ne isotopes it comes from the repulsive (attractive) neutron-neutron interaction between the 1$d_{5/2}$-2$s_{1/2}$ (1$d_{5/2}$-1$d_{5/2}$) orbitals as the neutron 1$d_{5/2}$ orbital is filled while the proton-neutron interaction is attractive for p(1$d_{5/2}$)-n(1$d_{5/2}$) leading to less configuration mixing, correspondingly smaller deformation, and a smaller point proton radius. For nitrogen isotopes the proton-neutron tensor interaction is more attractive for p(1$p_{1/2}$)-n(1$d_{5/2}$) orbitals, thereby reducing the gap between proton 1$p_{1/2}$ and 1$p_{3/2}$ orbitals when more neutrons are added in the $d_{5/2}$ orbital, hence resulting in a small point proton radius because of the lowering of the 1$p_{1/2}$ orbital. This attractive interaction also lowers the neutron 1$d_{5/2}$ orbital leading to the $N$ = 14 shell gap. It is therefore, reflected also in a dip in the matter radius and neutron skin thickness at $N$ = 14. 

\begin{figure}
\includegraphics[width=10cm, height=5cm]{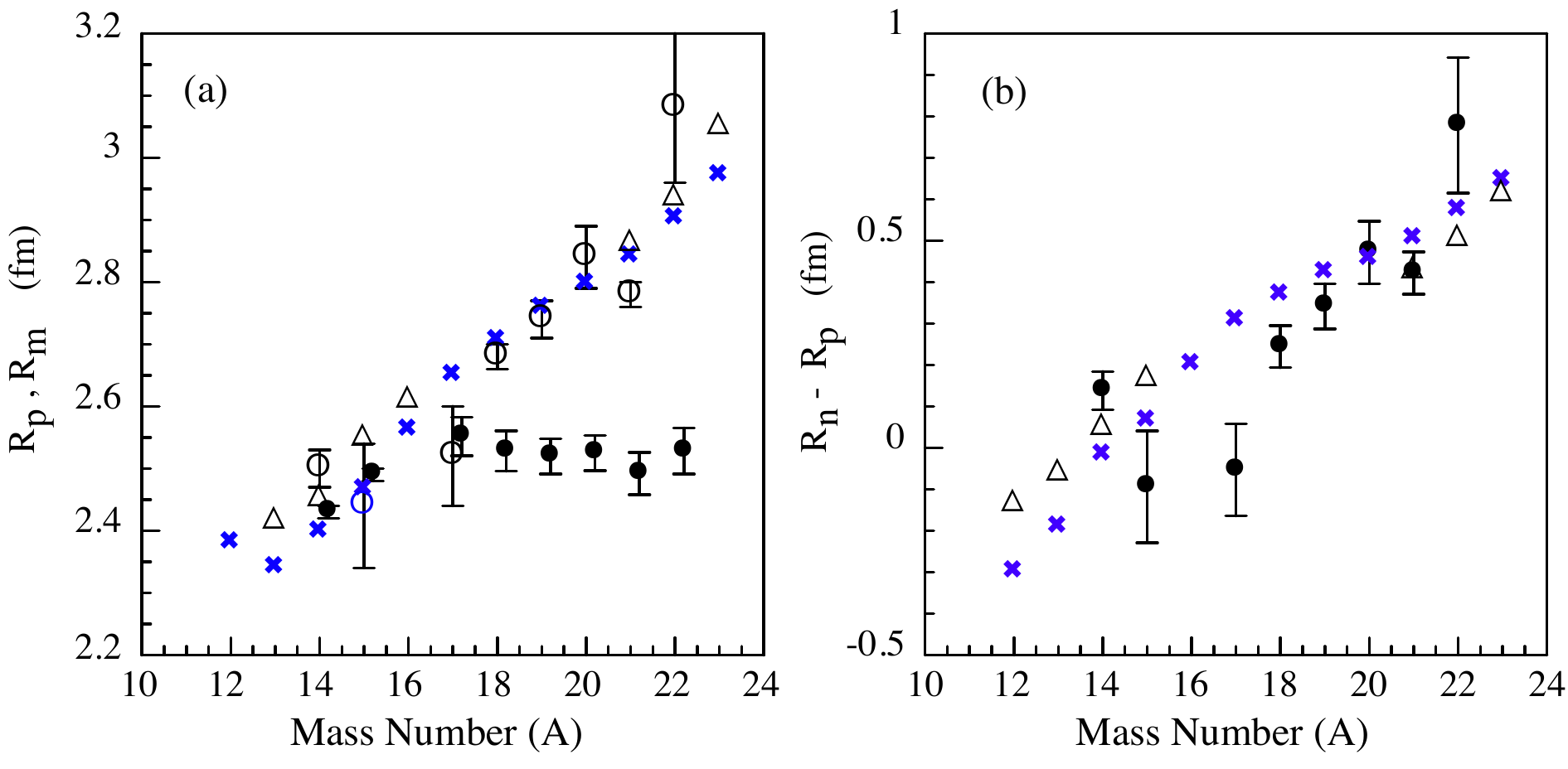}
\caption{\label{fig:epsart} (Color online) (a) The experimental point matter radii (black open circles) (b) measured neutron skin thicknesses (black filled circles) are compared to the respective theoretical calculations. The experimental point proton radii, R$_p^{ex,avg}$, are shown by black filled circles in (a) where those for $^{14,15}$N are from $e^-$ scattering. The blue crosses correspond to VS-IMSRG radii, with black open triangles are coupled-cluster radii computed with oxygen cores. }
\end{figure} 

In summary, the point proton radii for neutron-rich $^{17-22}$N were measured from charge changing cross sections on
a carbon target at $\approx$900$A$ MeV. A thick neutron skin for $^{19-21}$N, consistent with {\it ab initio} model predictions, is found while for $^{22}$N a neutron halo-like structure develops. The rapid increase of the matter radii deviating from the $A^{1/3}$ rule ($A$ being the mass number) demonstrates the need for studying the radii of these neutron-rich light nuclei. The point-proton radii decrease within uncertainties from $^{17}$N to $^{21}$N. It may reflect a transition from deformation towards sphericity at the $N$ = 14 shell closure. The increase of the point proton radius beyond $^{21}$N within uncertainties is due to the effect of the 2$s_{1/2}$ neutron in $^{22}$N.   The variance of the different models for R$_p$ show need for the rich information from the new data. Shell model calculations for the point proton radii are in agreement for $^{17-22}$N and RMF predictions are in agreement for $^{18-22}$N.  {\it Ab initio}  computations of the point proton radii agree well for the most neutron-rich isotopes but are unable to predict the complete evolutionary pattern. The new data therefore, provide rich grounds for further development of the {\it ab initio} theories and the chiral interactions. 

The authors are thankful for the support of the GSI accelerator staff
and the FRS technical staff for an efficient running of the
experiment. The support from NSERC, Canada for this work is gratefully
acknowledged. 
The support of the PR China government and Beihang university under the
Thousand Talent program is gratefully acknowledged.  The experiment work 
is partly supported by the grant-in-aid program of the Japanese
government under the contract number 23224008. This work is partly supported by 
Grants-in-Aid for Scientific Research (JP15K05090) of the JSPS of Japan.
supported by the Office of Nuclear Physics, U.S. Department of Energy
(Oak Ridge National Laboratory), DE-SC0008499 (NUCLEI SciDAC
collaboration), NERRSC Grant No.\ 491045-2011, and the Field Work
Proposal ERKBP57 at Oak Ridge National Laboratory.  Computer time was
provided by the Innovative and Novel Computational Impact on Theory
and Experiment (INCITE) program.  TRIUMF receives funding via a
contribution through the National Research Council Canada. This
research used resources of the Oak Ridge Leadership Computing Facility
located in the Oak Ridge National Laboratory, which is supported by
the Office of Science of the Department of Energy under Contract No.
DE-AC05-00OR22725, and used computational resources of the National
Center for Computational Sciences and the National Institute for
Computational Sciences. The authors thank B. Davids for a careful reading of the manuscript.



\end{document}